# K-doped Ba122 epitaxial thin film on MgO substrate by buffer engineering

Dongyi Qin[1,2], Kazumasa Iida[3,2], Zimeng Guo[4,2], Chao Wang[5], Hikaru Saito[6,2], Satoshi Hata[5,4,2], Michio Naito[1,2], and Akiyasu Yamamoto[1,2]

[1] *Department of Applied Physics, Tokyo University of Agriculture and Technology, Koganei, Tokyo 184-8588, Japan.*
[2] *JST CREST, Kawaguchi, Saitama 332-0012, Japan.*
[3] *Department of Electrical Electronic Engineering, Nihon University, Narashino, Chiba 275-8575, Japan.*
[4] *Interdisciplinary Graduate School of Engineering Sciences, Kyushu University, Kasuga, Fukuoka 816–8580, Japan.*
[5] *The Ultramicroscopy Research Center, Kyushu University, 744 Motooka, Nishi, Fukuoka 819–0395, Japan.*
[6] *Institute for Materials Chemistry and Engineering, Kyushu University, Kasuga, Fukuoka 816–8580, Japan.*

Molecular beam epitaxy of K-doped Ba122 ($Ba_{1-x}K_xFe_2As_2$) superconductor was realized on a MgO substrate. Microstructural observation revealed that the undoped Ba122 served as a perfect buffer layer for epitaxial growth of the K-doped Ba122. The film exhibited a high critical temperature of 39.8 K and a high critical current density of 3.9 MA/cm$^2$ at 4 K. The successful growth of epitaxial thin film will enable artificial single grain boundary on oxide bicrystal substrates and reveal the grain boundary transport nature of K-doped Ba122.

## I. Introduction

The iron-based superconductor,[1] K-doped Ba122 ($Ba_{1-x}K_xFe_2As_2$), with a high critical temperature $T_c$ = 38 K and a high upper critical field $\mu_0 H_{c2}$ ~ 100 T, is one of the most promising materials for high-field applications.[2–4] For practical applications, it is necessary to minimize the adverse effect of the weak-link at grain boundaries, which limits transport superconducting current.[5] For the past decade, research on polycrystalline K-doped Ba122 has been conducted and demonstrated the improvement of transport critical current density ($J_c$) of bulks,[6,7] wires, and tapes.[8,9] Recently, a high $J_c$ of $1.1 \times 10^5$ A/cm$^2$ at 4.2 K under a magnetic field of 10 T was achieved by densification and uniaxial grain texturing.[10] For further improvement of $J_c$, understanding the nature of grain boundaries is necessary.

The transport measurements over a single grain boundary are a clear-cut method in understanding the nature of the grain boundary. So far, in the Co- and P-doped Ba122, the transport properties across the grain boundary have been evaluated using bicrystal thin films.[11,12] The results have both proven their advantageous grain boundary transport properties, namely lower $J_c$ suppression with the grain boundary angle in comparison with the high-$T_c$ cuprates.[13] Differently from Co- and P-doped Ba122, however, the volatility of K provides a serious problem in the epitaxial growth of K-doped Ba122. Recently, we have reported the successful epitaxial growth of K-doped Ba122 by employing fluoride substrates (CaF$_2$, SrF$_2$, and BaF$_2$).[14]

The growth of K-doped Ba122 epitaxial thin films on oxide substrates is essential for bicrystal experiments. Since the discovery of the 122-type pnictide compound,[2] several groups have attempted growing K-doped Ba122 epitaxial thin films on oxide substrates. Lee *et al.* have attempted the growth by a two-step method; room-temperature growth of undoped Ba122 and high-temperature post-annealing with K lump.[15] Naito *et al.* have performed the growth by molecular-beam epitaxy at low temperature (~350°C).[16] As for the endmember KFe$_2$As$_2$ (K122), Hiramatsu *et al.* have prepared the film by high-temperature post-annealing of K-rich K122 film.[17,18] All three groups reported the *c*-axis oriented films, however, truly epitaxial (both *c*-axis oriented and in-plane aligned) film was only achieved in K122.[18] In this study, we employed MgO(001) as an oxide substrate, yet it turned out that K-doped Ba122 films directly grown on MgO(001) are not epitaxially grown.[14] To realize ideal epitaxy, proper buffer layers have to be introduced in between lower MgO substrate and upper K-doped Ba122 layers. We selected undoped Ba122 as a material for the buffer layer because it is capable of high-temperature growth, has perfect lattice matching with K-doped Ba122, and is compositionally similar to K-doped Ba122.[*]

The resultant film demonstrated a high $T_c$ of 39.8 K and a high $J_c$ of 3.9 MA/cm$^2$. The successful growth of epitaxial thin films on MgO substrates with a mega order of magnitude of $J_c$ is the first step toward elucidating the nature of the grain boundaries of K-doped Ba122.

## Experimental procedure

Film growth was performed using custom-design molecular-beam epitaxy (MBE) equipped with various element rate monitoring systems. The details of the deposition method are described in our previous report.[14] Briefly, all elements except for K were supplied from resistive heating of pure metal with a purity of 99+%, 99.98%, and 99.9999% for Ba, Fe, and As, respectively. Elemental K was supplied from In-K alloy (In$_8$K$_5$)[16,20,21] because of the ease of controlling evaporation rate and safety issues. Electron impact emission spectrometry (EIES) and atomic absorption spectrometry (AAS) were employed for the real-time rate monitoring of various elements: EIES for Ba and Fe, and AAS for K. K-doped Ba122 was grown at 400°C after the growth of the undoped Ba122 buffer layer at 720°C. The amount of supplied K was almost equal to the optimum (~0.40) or slightly lower level to obtain the highest $T_c$. The growth rate for both processes was ~1.5 Å/s and the deposition time was 2.5 min for the undoped layer and 10 min for the K-doped layer. In this study, we used MgO(001) substrates, which are commercially available in bicrystal form. K-doped Ba122 films grown on Ba122-buffered MgO substrate showed almost no appreciable degradation in the atmosphere.

---

[*] Previously, Haindl *et al.* have prepared SmFeAsO epitaxial thin film by introducing an undoped Ba122 buffer layer between upper SmFeAsO and lower MgO(001) substrate.[19]

The crystal structure of the films was evaluated by x-ray diffraction (XRD) and transmission electron microscopy (TEM). The cross-sectional TEM samples were prepared by focused ion beam (FIB) in a scanning electron microscope (SEM, Thermo Fisher Scientific Scios). The scanning-TEM (STEM, Thermo Fisher Scientific Titan G2 Cubed 60-300) was utilized for microstructural analysis equipped with bright-field (BF) and annular dark-field (ADF) detectors, and chemical compositional analysis by energy-dispersive x-ray spectroscopy (EDS). These analyses were operated at an acceleration voltage of 300 kV. The temperature dependence of in-plane resistivity was measured using a standard four-probe method. The resistivity of the film was calculated using the total thickness of the (Ba,K)122/Ba122 bilayer because the resistivity of the undoped Ba122 and K-doped Ba122 is comparable. Magnetic measurements were performed using on a rectangular slab (dimension of 1.512 mm × 2.194 mm × 80 nm and 2.091 mm × 2.540 mm × 80 nm for the temperature dependence of magnetic susceptibility and magnetic field dependence of $J_c$, respectively) a superconducting quantum interference device. The Bean critical-state model was applied to estimate $J_c$ values.[22]

## III. Results and discussion

Figure 1(a) shows the out-of-plane XRD pattern of K-doped Ba122 epitaxial thin film, which is deposited on the Ba122-buffered MgO substrate. Sharp (00$l$) ($l$ = 2, 4, 6, 8, 10) peaks with no misorientation were detected in the whole angle range. The (002) peak which is around $2\theta$ ~13.5° is a single peak, however, the split in the (00$l$) peaks appear above the (004) peak. With regard to the double peaks, the peak with higher intensity at a lower angle is considered to be the peak from the doped phase because the $c$-axis length is elongated with K doping in the K-doped Ba122 system. Accordingly, the separation of the peaks between the undoped Ba122 layer and the K-doped Ba122 layer enlarges at a higher angle region. The $c$-axis length of Ba122 and K-doped Ba122 was estimated to be 12.906Å and 13.188Å, respectively. It should be mentioned that the $c$-axis length of the K-doped Ba122 was much shorter than that of the $Ba_{0.6}K_{0.4}Fe_2As_2$ epitaxial thin films grown on $CaF_2$(001) substrate (13.380Å).[14]

Figure 1(b) shows the in-plane XRD pattern around the (103) reflection peak. Single domain fourfold symmetry with full width at half maximum (FWHM) of $\Delta\phi$~1.40° confirms that the film is in-plane aligned. In addition, the cube-on-cube epitaxial relationship between (Ba,K)122/Ba122 film and MgO substrate is confirmed. By comparing the crystallinity of the film grown on MgO substrate ($\Delta\theta_{008}$ = 0.66° and $\Delta\phi_{103}$ = 1.40°) with our previously reported film grown on $CaF_2$ substrate ($\Delta\theta_{008}$ = 0.52° and $\Delta\phi_{103}$ = 1.39°), it can be concluded that high crystallinity epitaxial K-doped Ba122 thin film was successfully grown on MgO substrate by introducing undoped Ba122 as a buffer layer. The $a$-axis length of K-doped Ba122 was estimated to be 3.931Å, which is longer than that of $Ba_{0.6}K_{0.4}Fe_2As_2$ epitaxial thin films grown on $CaF_2$(001) substrate (3.889Å).[14] The role of the Ba122 buffer layer is not clear at this stage. The possible reasons for the epitaxial growth are the improved lattice matching and the avoidance of direct contact between the K-doped Ba122 film and the oxygen in the oxide substrate due to the introduction of a buffer layer.

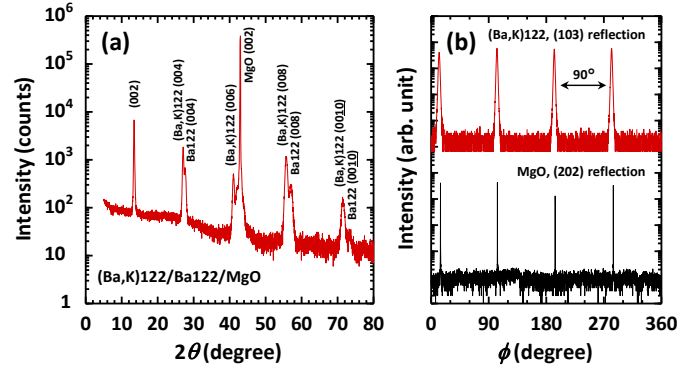

Figure 1. (a) Out-of-plane XRD pattern for K-doped Ba122 epitaxial thin film grown on Ba122-buffered MgO(001) substrate. The peaks of the (00$l$) are split by a small difference between the K-doped Ba122 epitaxial film and the undoped Ba122 buffer layer. (b) Comparison of the in-plane XRD patterns for the (103) reflection of K-doped Ba122 and the (202) reflection of MgO substrate.

In the cross-sectional BF-STEM image of the K-doped Ba122 film on MgO substrate, figure 2(a), thickness measurements were performed at several randomly selected locations for the Ba122 buffer layer and K-doped Ba122 layer, respectively. The statistics of the measured data yielded a thickness of 22 ± 2 nm for the Ba122 buffer layer and 80 ± 7 nm for the K-doped Ba122 layer. Columnar bright and dark contrasts in the epitaxial superconducting film suggest that the film contains many defects and columnar grain boundaries. The detailed atomic-resolution STEM analysis of these subgrain boundaries reveals that they are mostly low-angle grain boundaries, as shown in figure 2(b). Figure 2(d) and (e) show the atom resolution nanostructure of the interfaces between the K-doped Ba122 and the Ba122 layers, the Ba122 layer and MgO substrate, respectively. The perfect epitaxial relationship and interface between K-doped Ba122 and Ba122 buffer layer is revealed clearly in figure 2(d). From this image, the $c$-axis length of Ba122 and K-doped Ba122 are also estimated at 13.01Å and 13.22Å respectively (calibrated by $c_{MgO}$ = 4.212Å). The SAED pattern in figure 2(c) shows the epitaxial relationship between Ba122 and MgO substrate, that is (001)[100]-Ba122 || (001)[100]-MgO. The EDS analysis shown in figure 2(f) exhibits the changes of the chemical composition from undoped Ba122 to K-doped Ba122 mainly on barium and potassium. This result demonstrates that the interdiffusion of K is almost negligible, probably due to the low growth temperature (~400°C) of K-doped Ba122 films, which is consistent with the separation of the peaks observed in figure 1(a). The potassium concentration of the K-doped Ba122 film was estimated to be 36%, which is slightly lower than the optimal value of 40%.

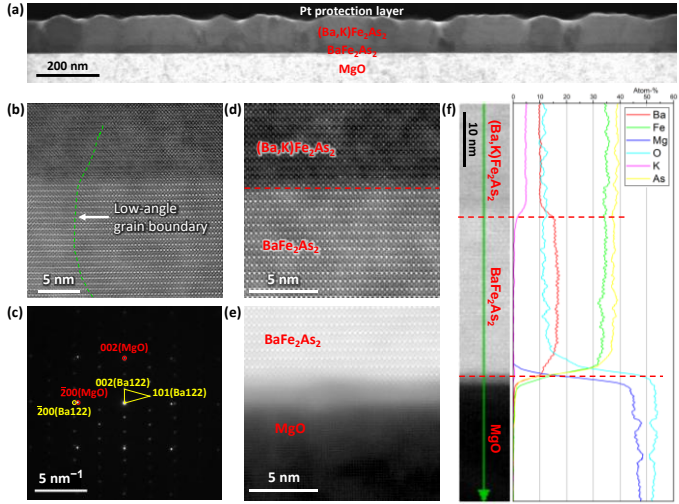

Figure 2. (a) Low magnification BF-STEM image of the K-doped Ba122 superconducting film on single-crystal MgO substrate with a buffer layer of undoped Ba122. (b), (d), and (e) show the atomic resolution ADF-STEM images corresponding to three feature regions: the low-angle grain boundary, the interface between K-doped Ba122 film and undoped Ba122 buffer layer, the interface between buffer layer and MgO substrate, respectively. (c) The select-area electron diffraction (SAED) pattern is according to the area which covers all three layers. (f) The line-scan EDS in scanning TEM mode, which shows the atomic percentages of each element in the K-doped Ba122, Ba122, and MgO layers.

Figure 3(a) shows the temperature dependence of resistivity of the film. The transition temperature is as high as 39.8 K with a sharp transition of $\Delta T_c = 0.9$ K. The observed $T_c$ is higher by ~3 K than $T_c$ in K-doped Ba122 epitaxial thin films grown on fluoride substrates.[14,23] The $T_c$ enhancement is considered to be due to the epitaxial strain (also/or thermal expansion mismatch): $a_0 = 3.917$Å for $Ba_{0.64}K_{0.36}Fe_2As_2$,[24] 3.863Å for $CaF_2$, and 4.212Å for MgO. The longer $a_0$ of MgO should introduce in-plane tensile and out-of-plane compressive strain. Normal-state resistivity at room temperature was 346 μΩcm, which should be compared with the values of single crystal (~310 μΩcm)[25] and polycrystalline bulk (~480 μΩcm)[26]. Residual resistivity ratio $RRR$ of 5.8 was smaller than that of single crystal (~11.6) and polycrystalline bulk (~6.8). Generally, $RRR$ increases with the improved crystallinity and reduced impurity concentration. In addition, $RRR$ varies with K content and strain induced by external pressure. Possible reasons for the relatively small $RRR$ are an electric conductivity carried in parallel by the undoped Ba122 buffer layer, low-angle grain boundaries, and epitaxial strain.

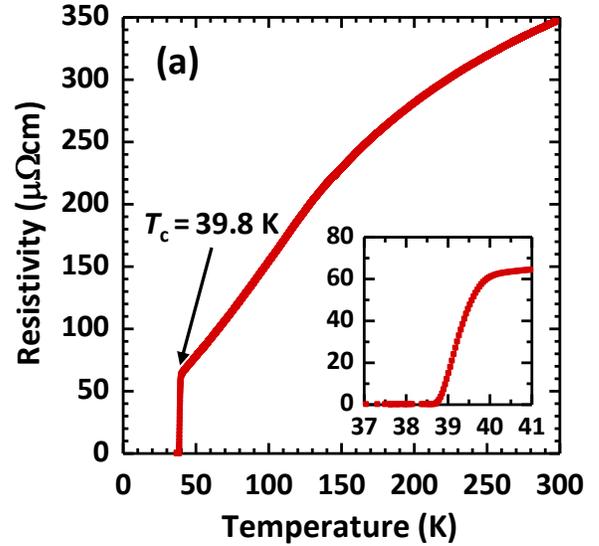

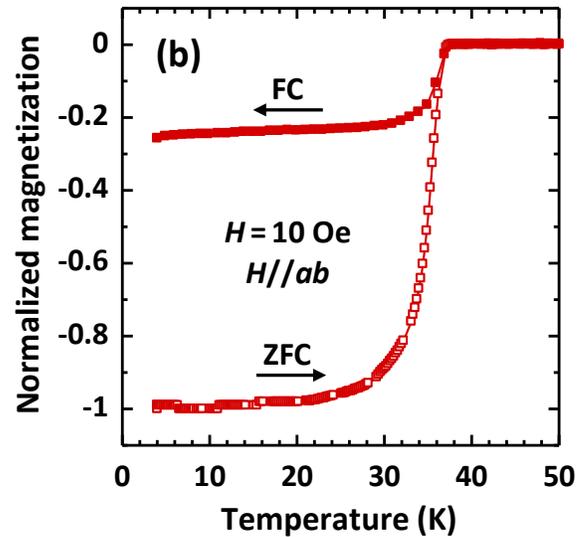

Figure 3. (a) Temperature dependence of in-plane resistivity for K-doped Ba122 epitaxial thin film on Ba122-buffered MgO substrate. Inset shows the temperature dependence of resistivity near $T_c$. (b) Temperature dependence of normalized magnetic susceptibility for the K-doped Ba122 epitaxial thin films grown on Ba122-buffered MgO substrate for the ZFC and FC process.

Figure 3(b) shows the temperature dependence of normalized magnetic susceptibility for both the zero-field-cooling (ZFC) process and the field-cooling (FC) processes. The data were normalized to their absolute value at 4 K. The diamagnetic signal was observed for the ZFC process below 37 K, which roughly follows the zero-resistance temperature observed on the transport measurement.

Figure 4 shows the magnetic field dependence of $J_c$ at different temperatures when a magnetic field is applied parallel to the $c$-axis direction for the K-doped Ba122 film grown on Ba122-buffered MgO. A reasonably high self-field $J_c$ of 3.9 MA/cm$^2$ was obtained at 4 K, exceeding the values reported for K-doped Ba122 single crystals.[27–29] The $J_c(\mu_0H)$ behavior of the film at 16 K is comparable to that of the film grown on $CaF_2(001)$ substrate at 4 K (black dashed line in figure 4). In comparison to the highest $J_c$ reported in Co-doped Ba122 epitaxial thin film (5.6 MA/cm$^2$),[30] the $J_c$ value is lower in self-field, but higher above 4 T. It is noteworthy that the $J_c$ value exceeds 1 MA/cm$^2$ under a high field of 7 T suggesting that the film has outstanding characteristics in magnetic fields.

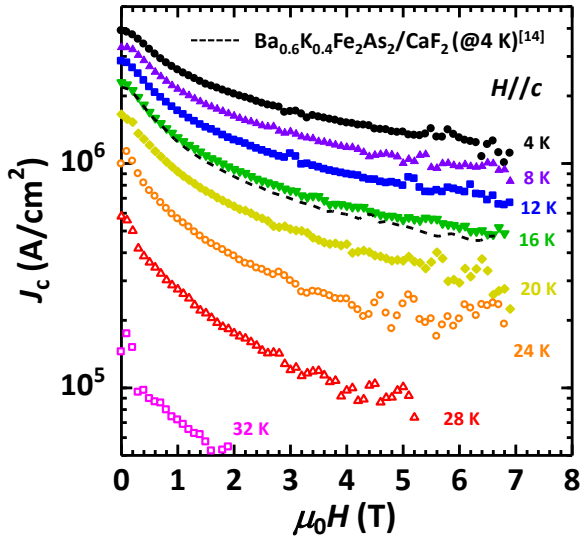

Figure 4. Magnetic field dependence of critical current density measured at different temperatures. The black dashed line is the 4 K $J_c(\mu_0 H)$ curve for $Ba_{0.6}K_{0.4}Fe_2As_2/CaF_2(001)$.[14]

## IV. Conclusion

In this study, we demonstrated the first successful epitaxial growth of K-doped Ba122 thin film on oxide substrate by employing undoped Ba122 as a buffer layer. The epitaxial growth was confirmed from both XRD and TEM observation. A high $T_c$ of 39.8 K exceeding the bulk value has been achieved, which is possibly due to epitaxial strain (and/or thermal expansion mismatch). A reasonably high self-field $J_c$ of 3.9 MA/cm$^2$ with $J_c > 1$ MA/cm$^2$ under a magnetic field of 7 T confirmed superior magnetic field dependence of K-doped Ba122.

## Acknowledgement

This work was partly supported by JST CREST (Grant No. JPMJCR18J4) and Advanced Characterization Platform of the Nanotechnology Platform Japan (Grants No. JPMXP09-A-19-KU-1003 and 1004) sponsored by the Ministry of Education, Culture, Sports, Science and Technology (MEXT), Japan.